      \documentstyle[12pt,agums]{article}

\lefthead{ANTIOCHOS AND DEVORE}

\righthead{HELICITY AND 3D MAGNETIC RECONNECTION SIMULATIONS}

\authoraddr{S. K. Antiochos,
Code 7675,
Naval Research Lab,
Washington, DC 20375-5352.
(e-mail: antiochos@nrl.navy.mil}

\authoraddr{C. R. DeVore,
Code 6440,
Naval Research Lab,
Washington, DC 20375.
(e-mail: devore@nrl.navy.mil)}

\begin{document}

%
%

\title{The Role of Helicity in Magnetic Reconnection:
3D Numerical Simulations}

%
%

\author{Spiro K. Antiochos and C. Richard DeVore}
\affil{Naval Research Laboratory, Washington, D. C.}

%
%

\begin{abstract}

We demonstrate that conservation of global helicity plays only a minor
role in determining the nature and consequences of magnetic
reconnection in the solar atmosphere. First, we show that observations
of the solar coronal magnetic field are in direct conflict with
Taylor's theory. Next, we present results from three-dimensional MHD
simulations of the shearing of bipolar and multi-polar coronal
magnetic fields by photospheric footpoint motions, and discuss the
implications of these results for Taylor's theory and for models of
solar activity. The key conclusion of this work is that significant
magnetic reconnection occurs only at very specific locations and,
hence, the Sun's magnetic field cannot relax completely down to the
minimum energy state predicted by conservation of global helicity.

\end{abstract}


\begin{article}
\section{INTRODUCTION}

Magnetic reconnection has long been invoked as the physical mechanism
underlying much of solar activity. For example, reconnection is
believed to be the process driving many of the observed dynamic solar
events ranging from spicules to the largest and most energetic
manifestations of solar activity, coronal mass ejections (CME) and
eruptive flares.  In spite of the long and intensive study of
reconnection in the solar atmosphere, the process is still not well
understood, especially in three dimensions. One of the main
difficulties in developing a comprehensive understanding is that
reconnection may take on different forms depending on the details of
the physical situation. Consequently, any theory that can provide some
unifying insight into the nature of reconnection would be of great
benefit to understanding many aspects of solar activity. This is the
compelling motivation behind studies of magnetic helicity. Since
magnetic helicity is believed to be conserved during reconnection in
general, the hope is that helicity conservation may allow one to
determine the final state of a reconnecting system without having to
calculate the detailed dynamics of the evolution.  Helicity
conservation may also be able to provide some valuable information on
the dynamics. In this paper we argue, however, that helicity plays a
negligible role in determining the evolution of reconnecting magnetic
fields in the Sun's corona. It should be emphasized that by the term
``helicity'',  we refer in this paper solely to the global relative
helicity [\markcite{{\it e.g., Berger}, 1985}], which defines a single
invariant. We are not referring to the helicity density which defines
an infinite set of possible invariants. Only the global helicity is
believed to be conserved during reconnection.

The basic theory for using helicity conservation to determine the
evolution of magneto-plasmas has been developed by Taylor [1986].  For
Taylor's theory to be applicable to the solar corona, three key
statements must be true.  First, the helicity (global) is conserved
during reconnection.  Our numerical simulations agree well with this
statement --- the higher the magnetic Reynolds number of the
simulation, the better the agreement. Second, helicity is the only
topological quantity that is generally conserved during reconnection.
We believe that this assumption is also true, but our simulations
cannot test it, because they all begin with a potential field in which
a simple shear or twist flow is imposed on the photospheric
boundary. There are no knots or disconnected flux in the coronal
field, and no braiding motions or higher-order topologies produced by
the boundary flows. Since the complete topology of our fields is
contained in the helicity density, it is unlikely that there are any
global topological invariants other than helicity available to be
conserved.

It appears, therefore, that the first two requirements for Taylor's
theory are valid for our simulations, and probably for the corona as
well.  The final requirement is that {\it complete reconnection}
occurs, {\it i.e.}, the reconnection continues until the magnetic
energy achieves its lowest possible state.  Note that this statement
does not say anything about helicity, it is actually a model for
reconnection.  Unfortunately, this statement is completely wrong for
our simulations and, we believe, also for the Sun.

The physical reason for the failure of complete reconnection in the
corona is that it requires the formation of numerous current sheets,
or sheet-like current structures. But we, and others, have found from
both 2.5D and 3D simulations that due to photospheric line-tying,
current sheets do not form easily in the corona [e.g., \markcite{{\it
Mikic, Schnack, and Van Hoven}, 1989}; \markcite{{\it Dahlburg,
Antiochos, and Zang}, 1991}; \markcite{{\it Karpen, Antiochos, and
DeVore}, 1990}].  It is instructive to note that the Taylor theory is
closely related to Parker's nonequilibrium theory for coronal heating
[\markcite{{\it Parker}, 1972; 1979}].  The nonequilibrium theory also
proposes that in a 3D system, current sheets will form spontaneously
throughout the coronal volume.  But, there have been numerous
simulations testing nonequilibrium [{\it e.g.}, \markcite{{\it Van
Ballegooijen}, 1985}; \markcite{{\it Mikic, Schnack, and Van Hoven},
1989}; \markcite{{\it Dahlburg, Antiochos, and Zang}, 1991}], and to
our knowledge, no simulation produces these current sheets.  This does
not mean that current sheets cannot form or that reconnection does not
occur in the corona. Many simulations find that current sheets readily
form at magnetic separatrices [{\it e.g.}, \markcite{{\it Karpen,
Antiochos, and DeVore}, 1995; 1996; 1998}], and intense current
concentrations do form at those locations where the photospheric
motions produce exponentially growing gradients in footpoint
displacements, in particular, at stagnation points of the flow [{\it
e.g.}, \markcite{{\it Van Ballegooijen}, 1986}; \markcite{{\it Mikic,
Schnack, and Van Hoven}, 1989}; \markcite{{\it Strauss}, 1993};
\markcite{{\it Antiochos and Dahlburg}, 1997}].  But since
reconnection occurs only at these very specific locations, it is far
from complete, and Taylor's theory cannot be used to determine either
the final state of the field or its evolution. We assert, therefore,
that while the global helicity is conserved, it plays little role in
determining the corona's dynamics and evolution.

This conclusion is also evident from observations.  The Taylor theory
would predict that the coronal field evolves towards a linear
force-free field. For an infinite system like the corona, the only
linear force-free field with finite energy is the field which is
current-free in any finite volume [{{\it Berger,} 1985}].  Therefore if the
theory held, the coronal field would evolve {\it via} reconnection to
the potential field, in which case there would be no need for CMEs or
eruptive flares.  It may be argued that the Taylor theory should not
be applied to the corona as a whole, since the helicity is not
uniquely defined for an infinite system.  But, in fact, the Taylor
prediction for an infinite system is completely sensible.  If
reconnection could proceed freely, indeed it would be energetically
favorable for the field to transfer all its shear and twist to the
outermost field lines that extend toward infinity, such as the
field lines at the poles.  By transferring all the shear/twist to the
longest field lines, the field conserves its helicity, but brings its
energy down to the potential field value. The only problem with this
type of evolution for the solar corona is that it is never observed.

One could argue, however, that a Taylor process may occur in some
small portion of the corona, such as an active region, in which case
the field should evolve to a linear force-free state inside this
bounded domain.  But this prediction also disagrees with
observations. The canonical result from vector magnetograms and from
H$_\alpha$ observations is that the field is strongly sheared near
photospheric polarity-reversal lines (``neutral'' lines), and
unsheared or weakly sheared far from these lines 
[{\it e.g.}, \markcite{{\it Gary et al.}, 1987}; \markcite{{\it Falconer
et al.}, 1997}]. (By shear we mean
that the field lines appear to be greatly stretched out along the
reversal line.)  We show below that such a shear distribution can
explain the formation of prominences/filaments, which lends strong
support to the observations. But this observed localization of the
shear is {\it not} compatible with a linear force-free field.

In order to demonstrate this point, consider a simple analytic
model for the field.  Take the active region to consist of a 2.5D linear
force-free field arcade:
\begin{equation}
\vec{B} = \nabla \times (A(y,z) \hat x) + B_x(y,z) \hat x.
\end{equation}
Since this field must satisfy,
$ \nabla \times \vec{B} = \lambda \vec{B}$,
where $\lambda$ is a constant, we find that $B_x = \lambda A$, and 
the force-free equation reduces to the 
usual Helmholtz form, $\nabla^2 A + \lambda^2 A = 0$. One possible
solution is:
\begin{equation}
A = \cos (k y) \exp (-\ell z),
\end{equation}
where the wavenumbers $k, \ell$, and $\lambda$ are related by,
$ \lambda ^2 = k^2 - \ell^2$.
We have chosen the form of the flux function in Equation (2)
so that it corresponds to a bipolar arcade with a photospheric polarity
reversal line at $y = 0$, and a width $k y = \pi $ (this 
periodic solution actually corresponds to an infinite set of arcades.)

If the wavelengths in the vertical and horizontal direction are equal,
$\ell = k$, then $\lambda = 0$, and the solution reduces to the
potential field. However, if the vertical wavelength becomes larger
than the horizontal one $\ell < k$ (we expect the force-free field to
inflate upward), then the solution corresponds to a field with finite
shear, $B_x \ne 0$. Assuming that our bipolar arcade is at disk
center, then the observed shear of the field at the photosphere would
be given by the angle, $\theta = \arctan (B_x / B_y)$. If the field is
potential then $B_x = 0$, which implies that $\theta = 0$, and the
field lines are perpendicular to the polarity reversal line (the $x$
axis).  For the nonpotential case we find from Equations (1) and (2)
that $B_y = dA /dz = -\ell A$.  Hence, $\theta = - \arctan (\lambda /
\ell)$. The shear is constant throughout the region rather than being
localized near the polarity-reversal line.  Although this result has
been derived for only one family of solutions, it seems likely to hold
true in general. A linear force-free field must has a constant ratio
of electric current magnitude to magnetic field magnitude, and hence
must have shear everywhere.  But a broad shear distribution is in
total disagreement with numerous observations of the solar field [{\it
e.g.}, \markcite{{\it Gary et al.}, 1987}; \markcite{{\it Schmieder et
al.}, 1996}].

We conclude, therefore, that complete reconnection does not occur even
in small regions of the corona, and that helicity conservation is of
limited usefulness for determining the structure and evolution of the
coronal field.  We verify this conclusion with large-scale 3D
numerical simulations in the following sections.  The goal of our
simulations is to understand the formation and eruption of solar
prominences and the accompanying CME, but as will be demonstrated
below, the simulations also address the issues of the role of helicity
conservation in magnetic reconnection and the applicability of the
Taylor theory to the corona.

\section{SIMULATIONS OF BIPOLAR FIELDS}

The first simulation concerns the formation of prominences. Solar
prominences or filaments consist of huge masses of cool ($\sim 10^4$
K), dense ($\sim 10^{11}$ cm$^{-3}$) material apparently floating high
up in the hot ($\sim 10^6$ K), tenuous ($\sim 10^{-9}$ cm$^{-3}$)
corona [{\it e.g.}, \markcite{{\it Priest}, 1989}].
Prominences reach heights of over $10^5$ km, which is approximately
three orders of magnitude greater than the gravitational scale height
of the cool material.  Hence, the most basic question concerning
prominences is the origin of their gravitational support. It must be
due to the magnetic field; the field lines in the corona must have
hammock-like geometry so that high-density plasma can be supported
stably in the hammock [\markcite{{\it Priest}, 1989}].

A characteristic feature of all prominences is that they form over
photospheric polarity-reversal lines which exhibit strong shear.
Since many prominences are also observed to be very long compared to
their width or height, 2.5D models for their magnetic structure (a
magnetic arcade) have usually been considered. Both numerical
simulations and analytic theory showed, however, that 2.5D models of a
sheared bipolar arcade cannot produce field lines with the necessary
dips to support prominence material [\markcite{{\it Klimchuk}, 1990};
\markcite{{\it Amari et al.}, 1991}]. This led many to consider more
complicated topologies involving multi-polar systems or topologies
with flux disconnected from the photosphere, the so-called flux ropes
[{\it e.g.}, \markcite{{\it Priest and Forbes}, 1990}; \markcite{{\it
van Ballegooijen and Martens}, 1990}].

We have shown, however, that the lack of dipped field lines is only an
artifact of assuming translational symmetry, and that a sheared 3D
bipolar field readily develops the correct geometry to support
prominences [\markcite{{\it Antiochos, Dahlburg, and Klimchuk}, 1994};
\markcite{{\it Antiochos}, 1995}]. Our previous results were based on
a 3D static equilibrium code that computed the force-free field in the
corona given the connectivity of the field lines at the photosphere.
Here we present results from recent fully time-dependent 3D
simulations of photospheric shearing of a bipolar field. Since we
include the dynamics, these simulations also address the issues of
current-sheet formation, reconnection, and eruption.

The code uses a highly-optimized parallel version of our 3D
flux-corrected transport algorithms to solve the ideal MHD equations
in a finite-volume representation. The code is thoroughly documented
and available on the WEB under the auspices of NASA's HPCC program
({\it see http://www.lcp.nrl.navy.mil/hpcc-ess/}). The computational
domain consists of the rectangular box, $-20 \le x \le 20, \; -4 \le y
\le 4, \; 0 \le z \le 8$. We use a fixed, but very large non-uniform
Cartesian mesh of $462 \times 150 \times 150$ points. The initial
magnetic field is that due to a point dipole located at (0,0,-2) and
oriented along the $y$-axis, so that the polarity reversal line at the
photospheric plane (z = 0) corresponds to the $x$-axis. As boundary
conditions, we impose line-tying with an assumed shear flow and no
flow-through conditions at the bottom, and zero gradient conditions on
all quantities at the sides and top.  The initial plasma density is
uniform, and we neglect the effects of both plasma pressure and
gravity in this simulation, corresponding to a zero beta
approximation. Note, however, that the plasma is fully compressible
and all Alfven waves are included in the calculation,

We shear the field by imposing a flow at the photosphere that is
localized near the polarity reversal line. The shear vanishes for
$\vert y \vert > 1$, and for $\vert y \vert \le 1$ it has the form:
\begin{equation}
V_x = (8 \pi/ \tau) \sin(\pi t/\tau) \sin(\pi y),
\end{equation}
where the time scale for achieving the maximum shear $\tau =100$.
Even though we performed the simulations on the latest architecture
massively-parallel machines, it is still not possible to use observed
solar values for the shear properties.  Our shear extends over roughly
half the width of the strong field region on the photosphere, 
wider than is observed, and the average shearing velocity is
approximately 10\% the Alfven speed in the strong field region,
rather than the 1\% typical of the Sun. But even with these
limitations, dipped field lines form readily in the corona.

Plate 1 shows the magnetic configuration halfway through the shearing,
at $t = \tau / 2$. It is evident that the strongly sheared field lines
have dipped central portions.  The dips form as a result of the
balance of forces between the increased magnetic pressure of the
low-lying sheared field lines and the increased tension of the
unsheared overlying field. Since the unsheared flux is strongest at
the center of the system, the downward tension force is strongest
there, producing a local minimum in the height of the sheared
flux. Also shown in the Plate is the half-maximum iso-surface of
electric current magnitude. As expected, the current is concentrated
where the gradient of the shear is largest, in the boundary between
the sheared and unsheared field.

The field of Plate 1 reproduces all the basic observed features of
prominences; hence, we conclude that the magnetic structure of solar
prominences and filaments is simply that of a sheared 3D field.  It is
tempting to conjecture that continued shearing of this field
eventually leads to eruption. This is the basic hypothesis of the
tether-cutting model, which proposes that reconnection of the sheared
field either with itself or with the unsheared flux destroys the force
balance between sheared and unsheared flux, thereby allowing the field
to erupt outward explosively [\markcite{{\it Sturrock}, 1989};
\markcite{{\it Moore and Roumeliotis}, 1992}].  Note that the
tether-cutting model is physically similar to the Taylor theory since
both hypothesize that reconnection transfers shear from inner to outer
field lines.

Our simulation, however, shows no evidence for either tether-cutting
or a Taylor process. We continued the shearing up to $t = 100$, twice
as far as shown in Plate 1. We then set the photospheric velocity to
zero, and let the system relax for another 100 Alfven times. The total
magnetic energy and helicity are shown in Figure 1.  The system
appears to achieve a stable equilibrium with negligible loss of either
energy or helicity. A key point is that the system appears stable,
even though some ``reconnection'' does occur. By $t = 80$ the imposed
boundary shear is so extreme that even with our large grid we cannot
resolve it numerically, which produces ``current-sheets'' in the
corona, in particular, the current structure seen in Plate 1.  As a
result, reconnection (or perhaps more appropriately diffusion) occurs,
and helical field lines begin to appear at this time. However, the
appearance of helical lines is confined to the regions of strongest
shear, and does not propagate outward as would be necessary for
tether-cutting or for a Taylor process. We conclude, therefore, that
our simulation rules out both tether-cutting and the Taylor theory as
viable models for the corona.

\section{SIMULATIONS OF MULTI-POLAR FIELDS}

There are two fundamental reasons for the lack of eruption in the
simulation described above.  First, line-tying inhibits magnetic
reconnection in a topologically smooth field such as a simple
bipole. Second, eruption of the low-lying sheared flux requires the
overlying unsheared field to open as well, but the Aly-Sturrock limit
implies that no closed configuration can have sufficient energy to
reach this open state [\markcite{{\it Aly}, 1984; 1991}; 
\markcite{{\it Sturrock}, 1991}].

We have argued that a multi-polar magnetic topology overcomes both
these problems, and have proposed a ``breakout'' model for prominence
eruptions and coronal mass ejections [\markcite{{\it Antiochos},
1998}; \markcite{{\it Antiochos, DeVore, and Klimchuk},
1999}]. Line-tying does not inhibit current-sheet formation at the
separatrix surfaces between flux systems, and these current sheets can
lead to sustained reconnection at separator lines [\markcite{{\it
Karpen, Antiochos, and DeVore}, 1995; 1996; 1998}]. Furthermore, a
multi-flux topology makes it possible to transfer the unsheared
overlying flux to neighboring flux systems, thereby allowing the
sheared field to erupt outward while keeping the unsheared flux
closed.  This allows the system to erupt explosively while still
satisfying the Aly-Sturrock energy limit [\markcite{{\it Antiochos,
DeVore, and Klimchuk}, 1999}].

We show below results from our first 3D simulation of the breakout
model.  The simulation domain in this case consists of the region: $-3
\le x \le 3, \; -3 \le y \le 3, \; 0 \le z \le 3$, with a fixed,
non-uniform grid of $166 \times 166 \times 86$ points. The initial
magnetic field is that due to three point dipoles: one located at (0,
25, -50), with magnitude unity, and pointing in the $+y$ direction;
another located at (0, 1, -1), with magnitude 10, and pointing in the
$-y$ direction; and the third at (0, 0, -0.5), with magnitude 10 and
pointing in the $-z$ direction.  The initial density and pressure were
chosen so that the average plasma beta near the base is less than 0.1.
(This simulation did not use the zero-beta approximation.)

Plate 2a. shows the initial potential field of the simulation. The
magnetic topology consists of four flux systems due to four distinct
polarity regions at the photosphere.  There are two toroidal
separatrix surfaces that define the boundaries of these flux systems,
and their intersection in the corona defines a separator line, along
which rapid reconnection can occur [\markcite{{\it Antiochos}, 1998}].
This configuration corresponds to a so-called delta-sunspot region.
We impose similar boundary conditions as in the previous simulation,
and apply a shearing motion localized near the circular
polarity-reversal line of the delta-spot located at the center of the
bottom plane. The shear is such that it produces a rotation of $\sim
2\pi$ over a time interval of 100 Alfven times.

Plate 2b. shows the effect of this shear.  All field lines shown in
Plates 2 and 2b are traced from exactly the same set of
positions at the photosphere.  It is evident that field lines with
footpoints near the polarity reversal line have been twisted through
almost a full rotation, (this sheared flux system corresponds to the
sheared prominence bipole of Plate 1).  A careful examination of the
field lines in 2b reveals that some of the unsheared delta-spot flux
overlying the sheared flux has become open, {\it i.e.}, it extends to
the top boundary of the simulation box rather than closing down over
the sheared field. This transfer occurs as a result of reconnection
between the delta-spot flux and the neighboring open flux system.

The effect of another $ 2\pi $ rotation is shown in Plate 2c. Almost
all the delta spot flux in now open. We show in Figure 2 the
total magnetic energy and helicity of the system, equivalent to
Figure 1.  Near the end of the second shearing phase there is
clearly a burst of impulsive energy release and helicity ejection
through the top of the system.  The helicity within the simulation
volume decreases much faster than the energy, because of the eruption.
Therefore, the evolution of our simulated corona is the exact opposite 
of a Taylor process!

Of course, this claim is somewhat overstated because we expect that,
in fact, the helicity of the whole system (including the erupted
field) stays constant, whereas the magnetic energy of the whole system
decreases. It is interesting to note, however, that since solar
telescopes observe only a finite coronal volume, eruptive opening of
the magnetic field in that volume implies that the observed helicity
decreases to zero, but the observed magnetic energy asymptotes to the
open field value, which is generally well above the potential field
value. In this sense, an anti-Taylor process is an appropriate
approximation for the evolution of coronal regions observed during an
eruption.

A key result of our 3D delta-spot simulation is that that even though
a great deal of reconnection occurs, it is far from complete.  The
reconnection is confined to the separator line between the sheared and
unsheared regions. Consequently, the field never approaches a linear
force-free field. Instead it evolves toward an open state in which the
currents are concentrated in a thin sheet.  This also is opposite to
what is expected for a Taylor process.

In summary, we conclude that while helicity is conserved and may well
be the only topological quantity that is generally conserved during
reconnection, the actual amount of reconnection in the Sun's corona is
determined by the detailed magnetic topology of the particular
region. Consequently, the global helicity by itself yields little
information on coronal evolution.


\acknowledgments
This work was supported in part by NASA and ONR.


%
%

\end{article}

\newpage

%
%

   %
   %
   %
   %
   %
   %

\begin{figure}
\figurenum{1}
	\caption{The total magnetic energy and magnetic helicity (relative) of 
a bipolar field that is sheared for 100 Alfven crossing times and then allowed 
to relax for another 100 Alfven times.
}
\end{figure}

\begin{figure}
\figurenum{2}
	\caption{Total energy and helicity of a delta-sunspot field that 
undergoes 2 shearing phases, each of duration 100 Alfven crossing times.
}
\end{figure}

%
%

   %

	\begin{plate} 
\platenum{1} 
\platewidth{41pc} 
\caption{Structure of a bipolar magnetic field that has been sheared
by footpoint motions at the photosphere, bottom plane in the Plate.
Contours of normal-magnetic-field magnitude are plotted on the bottom
plane.  Four field lines with footpoints in the
shear region and two field lines with
footpoints outside this region are shown. Also plotted is the iso-surface of
electric current magnitude at half-maximum.}
\end{plate}
	\begin{plate}
\platenum{2}
\platewidth{41pc}
\caption{Evolution of a delta-sunspot magnetic field that is 
sheared by footpoint motions. {\bf 2a.} The initial current-free
magnetic configuration. {\bf 2b.} The field lines after a shear of 
$\sim 2\pi$. {\bf 2c.} The field lines after a shear of $\sim 4 \pi$. 
}
\end{plate}

\end{document}